\documentclass[twocolumn,floatfix,superscriptaddress,amsmath,showkeys,aps,eqsecnum]{revtex4-1}
\usepackage{bm}
\usepackage{xcolor}
\usepackage{hyperref}
\usepackage{makeidx}
\usepackage{amsmath}
\usepackage{graphicx}
\usepackage{dcolumn}
\usepackage{amsfonts}
\usepackage{amssymb}
\usepackage{color}
\usepackage{epstopdf}

\begin{document}

\title{Finite temperature effects in quantum systems with competing scalar orders}

\author{Nei Lopes}
\email{nlsjunior12@gmail.com (N. Lopes)}
\affiliation{Centro Brasileiro de Pesquisas F\'{\i}sicas, Rua Dr. Xavier Sigaud 150, Urca, 22290-180, Rio de Janeiro , Brazil}
\author{Daniel G. Barci}
\affiliation{Departamento de F\'{\i}sica Te\'orica, Universidade do Estado do Rio de Janeiro, Rua S\~ao Francisco Xavier 524, 20550-013,
Rio de Janeiro, RJ, Brazil.}
\author{Mucio A. Continentino}
\affiliation{Centro Brasileiro de Pesquisas F\'{\i}sicas, Rua Dr. Xavier Sigaud 150, Urca, 22290-180, Rio de Janeiro , Brazil}

\date{\today}

\begin{abstract}

The study of the competition or coexistence of different ground states in many-body systems is an exciting and actual topic of research, both experimentally and theoretically. Quantum fluctuations of a given phase can suppress  or enhance another phase depending on the nature of the coupling between the order parameters,  their dynamics and the dimensionality of the system. The zero temperature phase diagrams of systems with competing scalar order parameters with quartic and bilinear coupling terms have been previously studied for the  cases of a zero temperature bicritical point and of coexisting orders. In this work, we apply  the  \textit{Matsubara summation} technique from finite temperature quantum field theory to introduce the effects of thermal fluctuations on the effective potential of these systems. This is essential to make contact with experiments. We consider  two and three-dimensional materials characterized by a Lorentz invariant quantum critical theory. We obtain that in both cases, thermal fluctuations lead to  weak first-order temperature phase transitions, at which coexisting phases arising from quantum corrections become unstable. We show that above this critical temperature, the system presents scaling behavior consistent with that approaching a quantum critical point. Below the transition the specific heat has a thermally activated contribution with a gap related to the size of the domains of the ordered phases.  We show that the critical temperature ($T_c$) in the coexistence region decreases as a function of the distance to the  zero temperature classical bicritical point. This indicates that at the fine tuned value of this transition,   the system attains the highest $T_c$ in the region of  coexistence. 
\end{abstract}

\maketitle

\section{Introduction}
\label{sec:Introduction}

In condensed matter physics, and more specifically in strongly correlated materials one can find several systems~\cite{Braithwaite,Sundar,Balatsky1,Balatsky2,Aoki,Arce-Gamboa,1,Steglich,2,2.1,Park,4,5,6,7,8} that at low temperatures exhibit different phases as a function of some fine tuned control parameter, such as pressure, doping or magnetic field~\cite{9,10}. 

Among these systems, the most prominent are those presenting exotic types of competing/coexisting orderings. For instance, pnictides~\cite{11,12,13,14,15} that display competing antiferromagnetic(AF)-superconductor(SC) orders separated by a first-order phase transition~\cite{1,16,17,18}. Moreover, we can point out SC-AF-structural transitions that appear in the iron-arsenide SC~\cite{16,19}, or even unconventional coexistence between magnetic and SC orders in some iron-based compounds~\cite{8}. In the same sense, U and Ce-based heavy fermions~\cite{Steglich}, as well as high-T$_c$ cuprate materials~\cite{4,5,7} also present unusual coexistence of magnetism and superconductivity~\cite{Braithwaite,Sundar,Aoki}. All these experimental findings are  ongoing research topics that do not have a fundamental theory to describe them yet.

Very recently\cite{35,35b}, we have studied the stability of  zero temperature ground states of systems with competing orders in the presence of quantum corrections. In this paper, in order to make contact with experiments, we study  the effects of thermal fluctuations in  two different situations: the case of a bicritical point and that where two phases coexist (see Fig.~\ref{fig1}).  The interplay between finite temperature and quantum fluctuations give rise to interesting effects and provides clear predictions for the expected behavior of these systems. This brings our previous results  to the new level of a testable theory.
 
For consistency,  we consider  two real scalar order parameters interacting through quartic, as well as bilinear coupling terms, in both three (3d) and two-dimensions (2d). The dynamics of the systems are characterized by  Lorentz invariant critical theories~\cite{36,37,38,32}, i.e., with linear dispersion relations~\cite{35,35b}. The effects of finite temperatures are introduced by means of the \textit{Matsubara summation} formalism from finite temperature quantum field theory~\cite{Matsubara,Kapusta,Abrikosov, zohar}.

Generically, depending on the symmetry, dynamics and dimensionality of the systems,  for both 2d and 3d systems, the zero temperature classical bicritical point (ZTCBP) (or ``mean field" quantum critical point) may be unstable to quantum corrections giving rise to coexisting phases~\cite{35,35b} at zero temperature. Here, we show that with increasing temperature these coexisting orders become unstable due to thermal fluctuations. The finite temperature transition at the critical temperature ($T_c$) is a {\it weak first-order transition}~\cite{Pikin,Fernandez,Millis}, characterized by a small latent heat compared to thermal fluctuations. Also the existence of a scaling regime above this transition makes  difficult to distinguish it experimentally  from a continuous transition.

For the fine tuned case of coexisting phases, our results also present a weak first-order temperature phase transition at $T_c$. However, distinctively from  the case of the bicritical point, in the coexisting phase there is the emergence of a new characteristic length scale in the system related to the distance to the zero temperature classical critical point (ZTCCP). 

We have obtained that the effective potential presents two different  behaviors  at finite temperatures. For $T >> T_c$ it scales as $T^{(d+z)/z}$,  for both cases above. In our case, we have fix the dynamic exponent  $z=1$, since we have considered that both order parameters are described by a Lorentz invariance dynamics~\cite{36,37,38,32,35,35b}. In this regime, the  specific heat  $C/T \propto T^{(d-z)/z}$. This behavior is characteristic of systems approaching a quantum critical point (QCP)~\cite{10}. All happens as if the system ignores the presence of the \textit{weak first-order transition} that occurs at a lower temperature.  On the other hand, in the coexistence region, for low temperatures $T << T_c$, we show that the specific heat has a thermally activated contribution~\cite{10}, with a gap that we  relate to the length scale of the domains present in this region.  Moreover, we also show that $T_c$ decreases as a function of the distance to the ZTCCP in the coexistence region. This implies that we  observe the highest $T_c$  at the fine tuned value where the mean field critical point would be located (before considering quantum fluctuations).

The paper is organized as follows: 
in section~\ref{sec:FineteTVeff}, we present a brief overview of our previous recent results~\cite{35,35b} for zero temperature and the motivation to take into account the effects of thermal fluctuations in competing orders systems. In section~\ref{sec:formalism}, we introduce the standard \textit{Matsubara summation} formalism  to include finite temperature effects on the effective potential. In section~\ref{sec:results}, we present our results for the effects of thermal fluctuations on the phase diagrams of systems with competing scalar orderings. We consider the cases the system is fine tuned to a zero temperature classical bicritical point or with two coexisting ground states.  Finally, in section \ref{sec:Conclusions}, we sum up and discuss our main results with their implications for experiments.

\section{Finite temperature effective potential for competing orders}
\label{sec:FineteTVeff}
Firstly, let us briefly summarize the results~\cite{35,35b} of  the zero temperature one-loop effective potential approach~\cite{32,33,34,35,35b,39}. We have considered quantum corrections to the mean-field phase diagrams of  systems with competing orders. We have studied  the case of two real scalar order parameters interacting by means of quartic, as well as bilinear coupling terms~\cite{35,35b}. The quantum nature of the fields is described by linear and quadratic dispersions, with dynamic exponents $z=1$ and $z=2$~\cite{37,35,35b}, respectively. 
We have investigated the stability of the ZTCCPs to quantum corrections for both 3d~\cite{35} and 2d~\cite{35b} systems in two different cases: bicritical point and coexistence region. We showed explicitly how the (in)stability of these points in systems with competing orders depends on symmetry, dynamics and dimensionality~\cite{35,35b}. 
An unavoidable condition to make contact with experiments is to introduce thermal fluctuations. This is the focus of the present paper.

\subsection{Matsubara summation formalism for the effective potential}
\label{sec:formalism}

In 2d and 3d systems, the zero temperature effective potential including quantum corrections that describes two competing orders ($\varphi_{1,2}$), with the same Lorentz invariant dynamics ($z=1$) and coupled by quartic ($\lambda_{12}$) and bilinear interactions ($\delta_{1,2}$) is given by~\cite{35,35b},
\begin{align}
\begin{array}{cc}
\Gamma^{(1)}=\frac{1}{2}\int \frac{d^{d}k}{(2 \pi)^{d}} \ln\left[\left(1+\frac{b_{1}}{k^{2}+r_{1}}\right)\left(1+\frac{b_{2}}{k^{2}+r_{2}}\right)\right] \\
\\
-\frac{\left(3\delta_{1}\varphi_{1}^{2}+3\delta_{2}\varphi_{2}^{2}+4\lambda_{12}\varphi_{1}\varphi_{2}\right)^{2}}{2(B^{2}-A^{2})}\int \frac{d^{d}k}{(2 \pi)^{d}} \left(\frac{1}{k^{2}+A^{2}}-\frac{1}{k^{2}+B^{2}}\right) + ct
\end{array}
\label{eq: num3.1}
\end{align}
where $k^2=\omega^2+q^2$ (Euclidean space), $\textit{ct}$ are the counterterms, introduced to renormalize the theory, making the physical observables \textit{cut-off} independent, and
\begin{align}
\begin{array}{cc}
b_{1}=12\lambda_{1}\varphi_{1}^{2}+2\lambda_{12}\varphi_{2}^{2}+6\delta_{1}\varphi_{1}\varphi_{2} ; \ A^{2}= r_{1}+b_{1} &
\\ 
\\
b_{2}=12\lambda_{2}\varphi_{2}^{2}+2\lambda_{12}\varphi_{1}^{2}+6\delta_{2}\varphi_{1}\varphi_{2} ; \ B^{2}=r_{2}+b_{2}
\label{eq: num3.2}
\end{array}
\end{align}
where $r_1$ and $r_2$ are the distances to the zero temperature critical points at which the order parameters $\varphi_{1}$ and $\varphi_{2}$ vanish, respectively. We also introduce two energy scales $\Delta_{1,2}=\sqrt{b_{1,2}+r_{1,2}}$ that will play a role when we consider different regimes below.

In order to extend the effective potential to include finite temperature effects, we apply the standard \textit{Matsubara summation}~\cite{Matsubara,Kapusta,Abrikosov,zohar} procedure and perform in Eq.~(\ref{eq: num3.1}) the usual identifications and replacements, i.e., 
\begin{equation}
\frac{1}{2}\int \frac{d^{d}k}{(2 \pi)^{d}} \Rightarrow \frac{1}{2}\frac{1}{(2 \pi)^{d}}\int d^{d-1}p \  \left(T \sum_{\omega_n} \right)
\label{eq: num3.3}
\end{equation}
where, $d^{d}k = S_d k^{d-1} dk$ with $S_d = (2\pi)^{d/2}/\Gamma(d/2)$ and 
$\omega_n=2\pi n T$, 
with $n \in \mathcal{Z}$.

The \textit{Matsubara frequencies} ($\omega_n$)  are discrete, due to the periodic boundary conditions, $\varphi_i(0, x) = \varphi_i(\beta, x)$, and the fields should satisfy on the finite imaginary time axis $0\le \tau \le  \beta=1/T$ (we consider the Boltzman constant $k_B=1$ all along the paper).

It's worth to emphasize that  temperature effects do not introduce new divergences in the theory. For this reason, the only counterterms, needed to renormalize the theory, are those  introduced at zero temperature.  Finally, notice that the dimensionality is only contained in the \textit{momentum} integrations. For instance, in 2d we get the finite temperature effective potential,
\begin{align}
&\Gamma^{(1)} = V_{{eff}_{(T=0)}}^{(2+1)dim} \label{eq: num3.5} \\ 
+&\frac{1}{2}\frac{1}{(2 \pi)^{2}} \int dp \ p \Bigg\{ 2 T \ \ln\left[\frac{\left(1-e^{-\frac{\epsilon_{p_{1}}}{T}}\right)}{\left(1-e^{-\frac{\epsilon_{p_{2}}}{T}}\right)}\frac{\left(1-e^{-\frac{\epsilon_{p_{3}}}{T}}\right)}{\left(1-e^{-\frac{\epsilon_{p_{4}}}{T}}\right)}\right]
\nonumber   \\
-&\frac{\left(3\delta_{1}\varphi_{1}^{2}+3\delta_{2}\varphi_{2}^{2}+4\lambda_{12}\varphi_{1}\varphi_{2}\right)^{2}}{(B^{2}-A^{2})}
\!\!\left[\frac{f_{BE}(\epsilon_{p_{5}})}{\epsilon_{p_{5}}}-\frac{f_{BE}(\epsilon_{p_{6}})}{\epsilon_{p_{6}}}\right]\!\!\Bigg\}
\nonumber 
\end{align}
where $V_{{eff}_{(T=0)}}^{(2+1)dim}$ is the zero temperature one-loop effective potential~\cite{35b}, $\epsilon_{p_{1,3}}^{2}=p^2+r_{1,2}+b_{1,2}$, $\epsilon_{p_{2,4}}^{2}=p^2+r_{1,2}$, $\epsilon_{p_{5,6}}^{2} = p^2+(A,B)^2$, with $b_{1,2}$ and $(A,B)^2$ given in Eq.~(\ref{eq: num3.2}), and
\begin{equation}
f_{BE}(\epsilon_{p})=\frac{1}{e^{\epsilon_{p}/T}-1}
\label{eq: num26}
\end{equation}
is the Bose-Einstein distribution function~\cite{Matsubara,Kapusta,Abrikosov}.

The 3d expression is very similar to Eq. Eq.~(\ref{eq: num3.5}). The only difference is that the \textit{momentum} integration $\int dp\;  p$ should be replaced by $\int dp\;  p^2$. The explicit expression of the zero temperature contribution   $V_{{eff}_{(T=0)}}^{(3+1)dim}$ was computed in Ref. \onlinecite{35}.

\section{Finite temperature effects}
\label{sec:results}
To analyze the effects of finite temperature on the phase diagrams we need to solve the \textit{momentum} integral in Eq.~(\ref{eq: num3.5}).
We will do it in both cases, i.e., bicritical point and coexistence region. When it is not possible to get some analytical expression, we will solve the integrals numerically.

\subsection{Finite temperature effects in bicritical point for two-dimensional case: Preamble}
Let us firstly  recall the expression  for $V_{{eff}_{(T=0)}}^{(2+1)dim}$  in Eq.~(\ref{eq: num3.5}), in the case of a bicritical point~\cite{35b}.
\begin{align}
\begin{array}{cc}
\small{V_{{eff}_{(T=0)}}^{(2+1)dim}}=\underbrace{\lambda_{1}\varphi_{1}^{4}+\lambda_{2}\varphi_{2}^{4}+\lambda_{12} \varphi_{1}^{2} \varphi_{2}^{2}+\delta_{1} \varphi_{1}^{3}\varphi_{2}+\delta_{2} \varphi_{1}\varphi_{2}^{3}}_{classical \ term} \\
\\
\underbrace{-\frac{\sqrt{2}}{2 \pi}\left[\frac{1}{3}\left(b_{1}^{3/2}+b_{2}^{3/2}\right)+\frac{\left(3\delta_{1}\varphi_{1}^{2}+3\delta_{2}\varphi_{2}^{2}+4\lambda_{12}\varphi_{1}\varphi_{2}\right)^{2}}{4(\sqrt{b_{2}}+\sqrt{b_{1}})}\right]}_{quantum \ corrections}
\end{array}
\label{eq: num4.1}
\end{align}
where, $b_{1,2}$ are given by Eq.~(\ref{eq: num3.2}).

\begin{figure}[!t]
\begin{center}
\includegraphics[width=1\columnwidth]{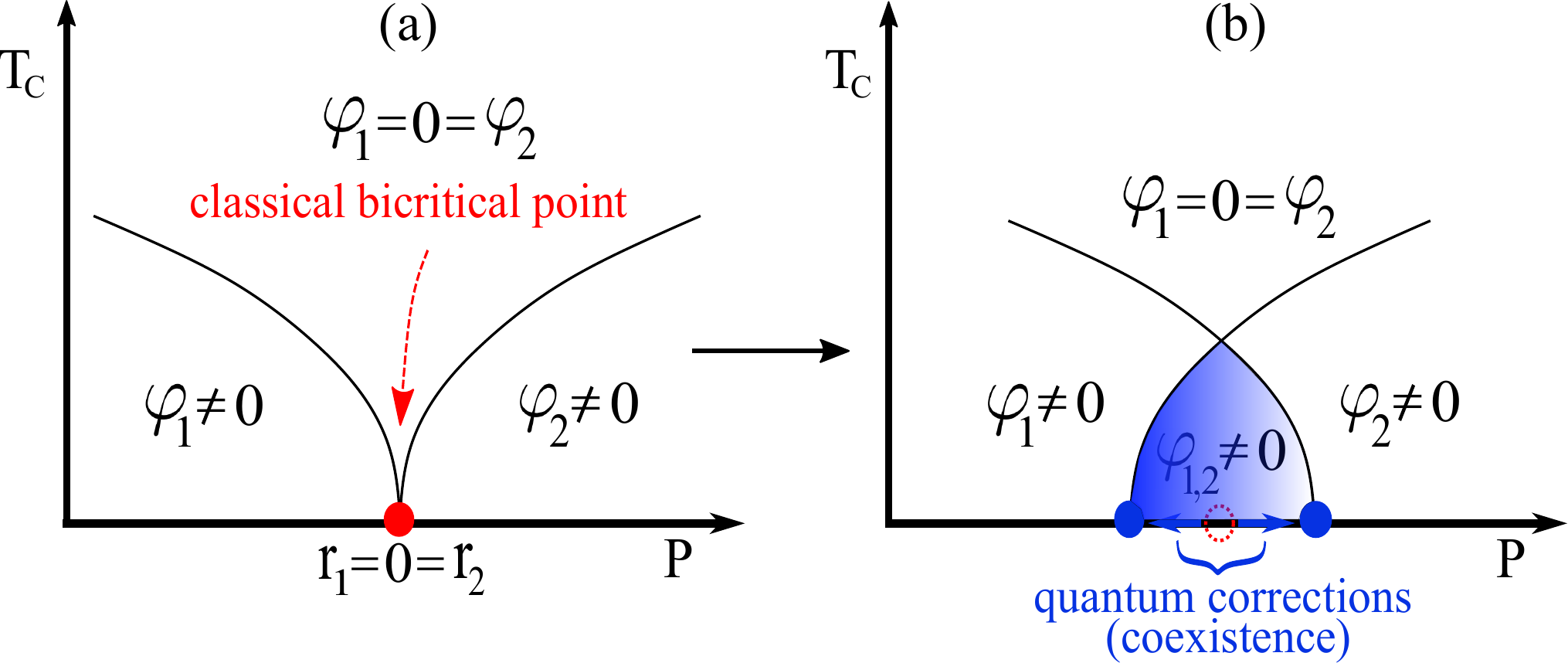} 
\end{center}
\caption{(Color online) The ZTCBP (red) is unstable to quantum corrections in 2d for quartic and bilinear couplings~\cite{35b}. In 3d it is stable for a quartic coupling between the fields  but unstable to a bilinear interaction~\cite{35}. The dynamics of the fields is described by  Lorentz invariant critical theories and give rise, in the unstable cases, to a region of coexistence ($\varphi_{1,2} \neq 0$ (shaded (blue) region)), as shown schematically in the phase diagram.}\label{fig1}

\end{figure}

From Eq.~(\ref{eq: num4.1}), we can see that  quantum corrections in 2d give rise to coexistence for any finite couplings, i.e., for quartic as well as bilinear interactions~\cite{35b} (see Fig.~\ref{fig1}). Conversely, in  the 3d case, only the bilinear coupling gives rise to coexistence~\cite{35}. 
Below, we  discuss the effects of finite temperature at a bicritical point.
\subsection{Finite temperature effects in bicritical point for two-dimensional case: Results}

We consider now the effects of finite temperature on the ZTCBP at $r_1=r_2=0$ in 2d systems. These parameters appear in definitions of $\epsilon_{p_{i}}$ in Eq.~(\ref{eq: num3.5}). It turns out in this case that it is not possible to obtain an analytical expression for the full temperature dependent effective potential, Eq.~(\ref{eq: num3.5}), and we have to analyze particular limits.

First, we consider the limit of low temperatures that correspond to $T << \Delta_{1,2}$. In this limit, one can neglect the last two terms depending on temperature in Eq.~(\ref{eq: num3.5}), since the first one dominates at this  temperature regime. We get the following equation for the effective potential,
\begin{align}
&V_{eff} = V_{{eff}_{(T=0)}}^{(2+1)dim}
\label{eq: num4.2} \\
+&\frac{1}{2}\frac{1}{(2 \pi)^{2}} \int dp \ p \left\{ 2 T \ \ln\left[\frac{\left(1-e^{-\frac{\epsilon_{p_{1}}}{T}}\right)}{\left(1-e^{-\frac{\epsilon_{p_{2}}}{T}}\right)}\frac{\left(1-e^{-\frac{\epsilon_{p_{3}}}{T}}\right)}{\left(1-e^{-\frac{\epsilon_{p_{4}}}{T}}\right)}\right]\right\}
\nonumber 
\end{align}
where, again, the term  $V_{{eff}_{(T=0)}}^{(2+1)dim}$  is given by Eq.~(\ref{eq: num4.1}).
In this equation for $r_{1,2}=0$, the terms involving $\epsilon_{p_{2,4}}$ have to be treated differently from those depending on $\epsilon_{p_{1,3}}$.
The former yields,
\begin{equation}
\int_0^{\infty} dp \ p \ \ln\left(1-e^{-\frac{\epsilon_{p_{2,4}}}{T}}\right) = - T^2  \zeta(3)  \approx  - 1.2  \hspace{0.1cm} T^{2}
\label{eq: num4.25}
\end{equation}
The latter, in the low temperature regime $T << \Delta_{1,2}$, yields from Eq.~(\ref{eq: num4.2}), 
\begin{equation}
\int_0^{\infty} dp \ p \ \ln\left(1-e^{-\frac{\epsilon_{p_{i}}}{T}}\right) \approx - \int p \ dp \ e^{-\frac{\epsilon_{p_{i}}}{T}}
\label{eq: num4.3}
\end{equation}
where we have used  that $\ln(1-x)\approx -x$.  From Eq.~(\ref{eq: num4.3}), we finally get,
\begin{equation}
- \int_0^{\infty} dp \ p \ e^{-\frac{\epsilon_{p_{1,3}}}{T}}=-T^2\left[1+\frac{\sqrt{b_{1,2}}}{T}\right]e^{-\frac{\sqrt{b_{1,2}}}{T}}.
\label{eq: num4.4}
\end{equation}

Therefore, we can rewrite Eq.~(\ref{eq: num4.2}), with the help of Eq.~(\ref{eq: num4.4}), Eq.~(\ref{eq: num4.25}) and Eq.~(\ref{eq: num4.1}) as follows
\begin{align}
\begin{array}{ccc}
V_{eff}=\underbrace{\lambda_{1}\varphi_{1}^{4}+\lambda_{2}\varphi_{2}^{4}+\lambda_{12} \varphi_{1}^{2} \varphi_{2}^{2}+\delta_{1} \varphi_{1}^{3}\varphi_{2}+\delta_{2} \varphi_{1}\varphi_{2}^{3}}_{classical \ term} \\
\underbrace{-\frac{\sqrt{2}}{2 \pi}\left[\frac{1}{3}\left(b_{1}^{3/2}+b_{2}^{3/2}\right)+\frac{\left(3\delta_{1}\varphi_{1}^{2}+3\delta_{2}\varphi_{2}^{2}+4\lambda_{12}\varphi_{1}\varphi_{2}\right)^{2}}{4(\sqrt{b_{2}}+\sqrt{b_{1}})}\right]}_{quantum \ corrections} \\
\underbrace{+\frac{T^3}{(2 \pi)^2}\left\{-\sum_{i=1}^{2}\left[1+\frac{\sqrt{b_{i}}}{T}\right]e^{-\frac{\sqrt{b_{i}}}{T}}+2.4\right\}}_{finite \ temperature \ effects}
\end{array}
\label{eq: num4.6}
\end{align}
where, $b_{1,2}$ are given by Eq.~(\ref{eq: num3.2}).

\begin{figure}[t]
\begin{center}
\includegraphics[width=1\columnwidth]{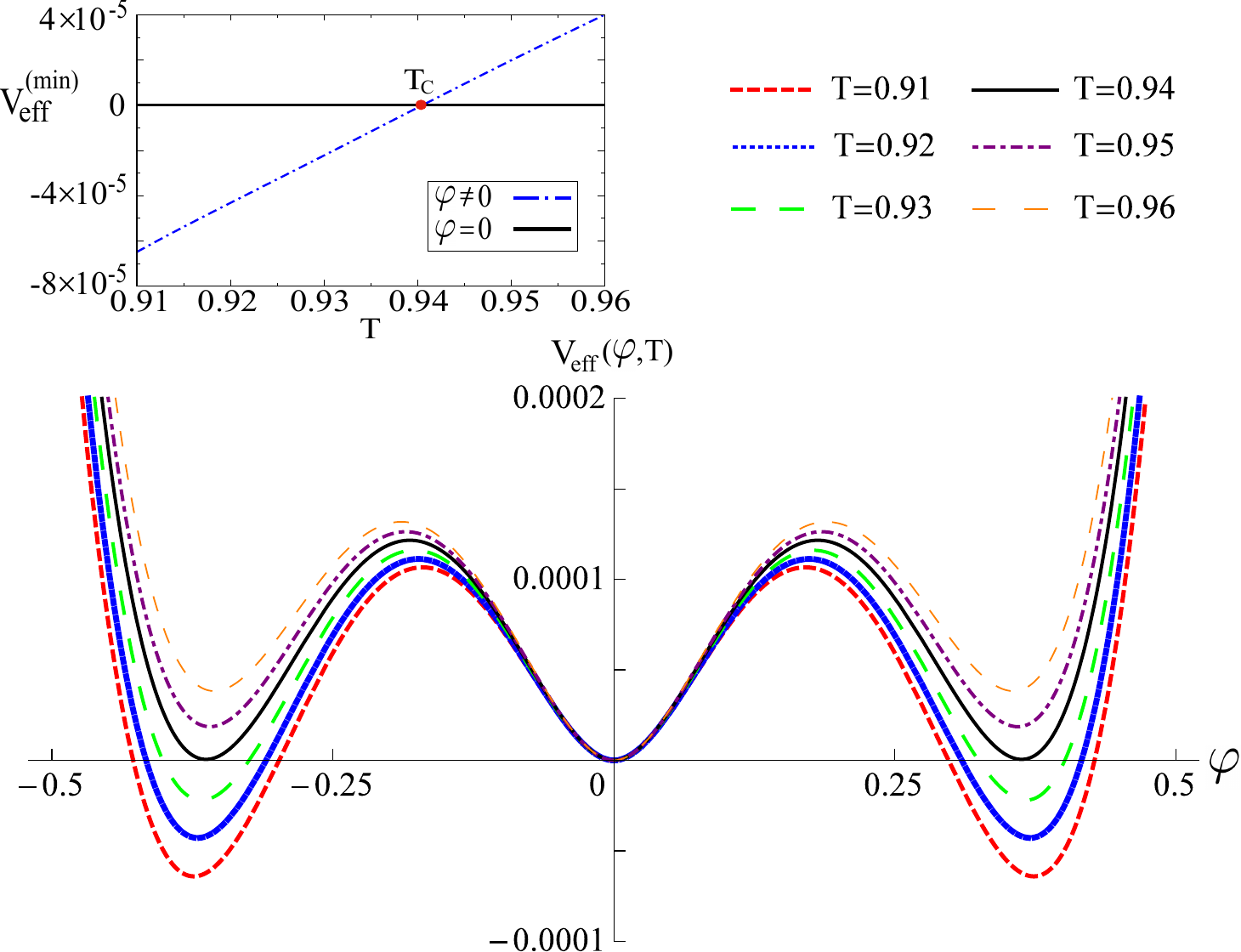} 
\caption{(Color online) The effective potential for a bicritical point in 2d systems with $z=1$ for different values of temperatures (Eq.~(\ref{eq: num4.6})).  Without loss of generality, we take $\lambda_1=\lambda_2=0.05$, $\lambda_{12}=0.01$, $\delta_1=\delta_2=0$, and $\varphi_1=\varphi_2=\varphi$. One can clearly see the signature of a  first-order temperature phase transition (solid (black) line) at $T_c \approx 0.94$,  when the \textit{minima} become degenerate. The tangent of the angle between the crossing energy lines (upper panel) is related to the latent heat L (see text).}\label{fig2}
\end{center}
\end{figure}

Initially, we investigate the effects of finite temperature plotting the effective potential for the bicritical case with $z=1$, that is, Eq.~(\ref{eq: num4.6}), for different values of temperature. Without loss of generality, in numerical calculations we take $\lambda_1=\lambda_2=0.05$ and $\lambda_{12}=0.01$ in energy units  to analyze the effects of the interaction term. For simplicity, one can also take $\delta_1=\delta_2=0$ in Eq.~(\ref{eq: num4.6}) for the two-dimensional case, since both couplings give rise to the same behavior~\cite{35b}. Furthermore,  we can take $\varphi_1=\varphi_2=\varphi$ to generate a two-dimensional plot since both order parameters are finite in the coexistence region, as shown in Fig.~\ref{fig2}. 

The coexistence phase induced by quantum corrections~\cite{35b} becomes unstable as we increase temperature, i.e., thermal fluctuations tend to disorder the system, as expected. This is shown in Fig.~\ref{fig2} (lower panel) where thermal fluctuations give rise to a \textit{weak first-order temperature phase transition}~\cite{Pikin,Fernandez,Millis} when the energy of the minima become degenerate (solid black line). 
This transition is associated with the  interchange of stability of the two phases, ordered and disordered, as can be also seen in Fig.~\ref{fig2} (upper panel). The tangent of the angle between the crossing energy lines is related to the latent heat $L$~\cite{10}. Using the numerical values of Fig.~\ref{fig2} we obtain $L/T_c \approx 0.0021 << 1$, consistent with the \textit{weak first-order} nature of this transition, i.e., this kind of transition presents a small latent heat compared to thermal fluctuations~\cite{10}, making it difficult to be experimentally distinguished from a continuous transition.

If we tune  the external parameter at the ZTCBP (see Fig.~\ref{fig1}~(a)), whereupon quantum corrections give rise to a coexistence region~\cite{35b} (see Fig.~\ref{fig1}~(b)), we can study the effects of  the temperature on the thermodynamic properties of the system. 

For $T >> T_c$, we obtain, from Eq.~(\ref{eq: num4.6}), a cubic temperature dependence in the effective potential for the two-dimensional case, in agreement with the expected  scaling form of the  free energy~\cite{10}, i.e., $f \propto T^{(d+z)/z}$, for a system approaching a Lorentz invariant QCP in d-dimensions.
\begin{figure}[t]
\begin{center}
\includegraphics[width=1\columnwidth]{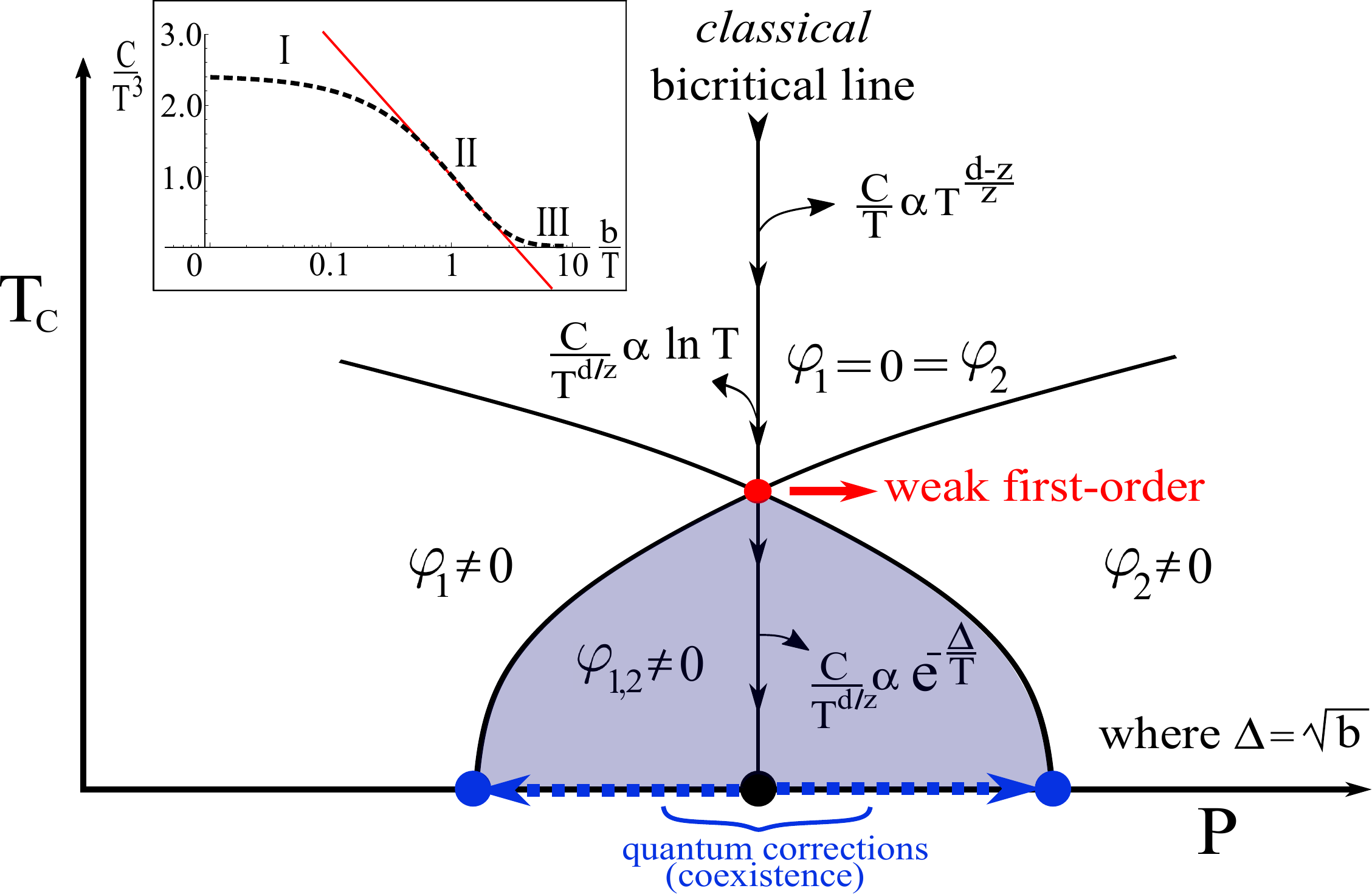} 
\caption{(Color online) Schematic phase diagram for a bicritical point in 2d systems with $z=1$ showing the different  scaling regimes of the specific heat.  Quantum corrections (blue arrows) give rise to a coexistence region ($\varphi_{1,2}\neq0$ (shaded gray)). At the classical bicritical line (see Fig.~\ref{fig1}), we can identify three different scaling regimes for the specific heat. At the temperature $T_c$ where both critical lines cross  (red) there is a \textit{weak first-order temperature phase transition}. Below  $T_c$,  in the coexistence region, there are gaps for thermal excitations, that is, $\Delta_{1,2}=\sqrt{b_{1,2}}$, where $b_i$ is given by Eq.~(\ref{eq: num3.2}). The inset shows the numerical results for the specific heat calculated from the temperature dependent integral in Eq.~(\ref{eq: num4.2}) for $T \gtrsim T_c$. The intermediate regime, i.e., regime II, is distinguished by a logarithmic dependence (solid red line in the  inset).}\label{fig3}
\end{center}
\end{figure}
This implies that the specific heat scales as $C/T \propto T^{(d-z)/z}$~\cite{10}. In fact, it presents two regimes for $T\gtrsim T_c$, as shown in Fig.~\ref{fig3}. A high temperature one, for $T >> T_c$, where $C/T \propto T^{(d-z)/z}$ and the system appears to be unaware of both the existence of a weak first-order transition at lower temperatures and of a coexistence phase for $T<T_c$. In this regime it behaves as approaching a QCP in d-dimensions with dynamic exponent $z$. 
At lower temperatures, but still for $T\gtrsim T_c$, the specific heat crosses over to a less universal regime, where it behaves as $C/T^{d/z} \propto -\ln T$, see Fig.~\ref{fig3}.  

At very low temperatures, $T << T_c$, the specific heat vanishes exponentially as a function of temperature, i.e., $C/T^{d/z} \propto \exp(-\Delta/T)$~\cite{10} as can be seen from Eq.~(\ref{eq: num4.6}). In this regime,  quantum corrections give rise to a coexistence region and there are gaps,  $\Delta_{1,2}=\sqrt{b_{1,2}}$ for thermal excitations in finite size domains, see Eq.~(\ref{eq: num4.4}). Then, we can define two new length scales  $\xi_{1,2}=1/\sqrt{2 b_{1,2}}$, that are essentially associated with the size of  domains that begin to form once we reach the coexistence region cooling down the system~\cite{10}, see arrows in Fig.~\ref{fig3}. 

\begin{figure}[t]
\begin{center}
\includegraphics[width=1\columnwidth]{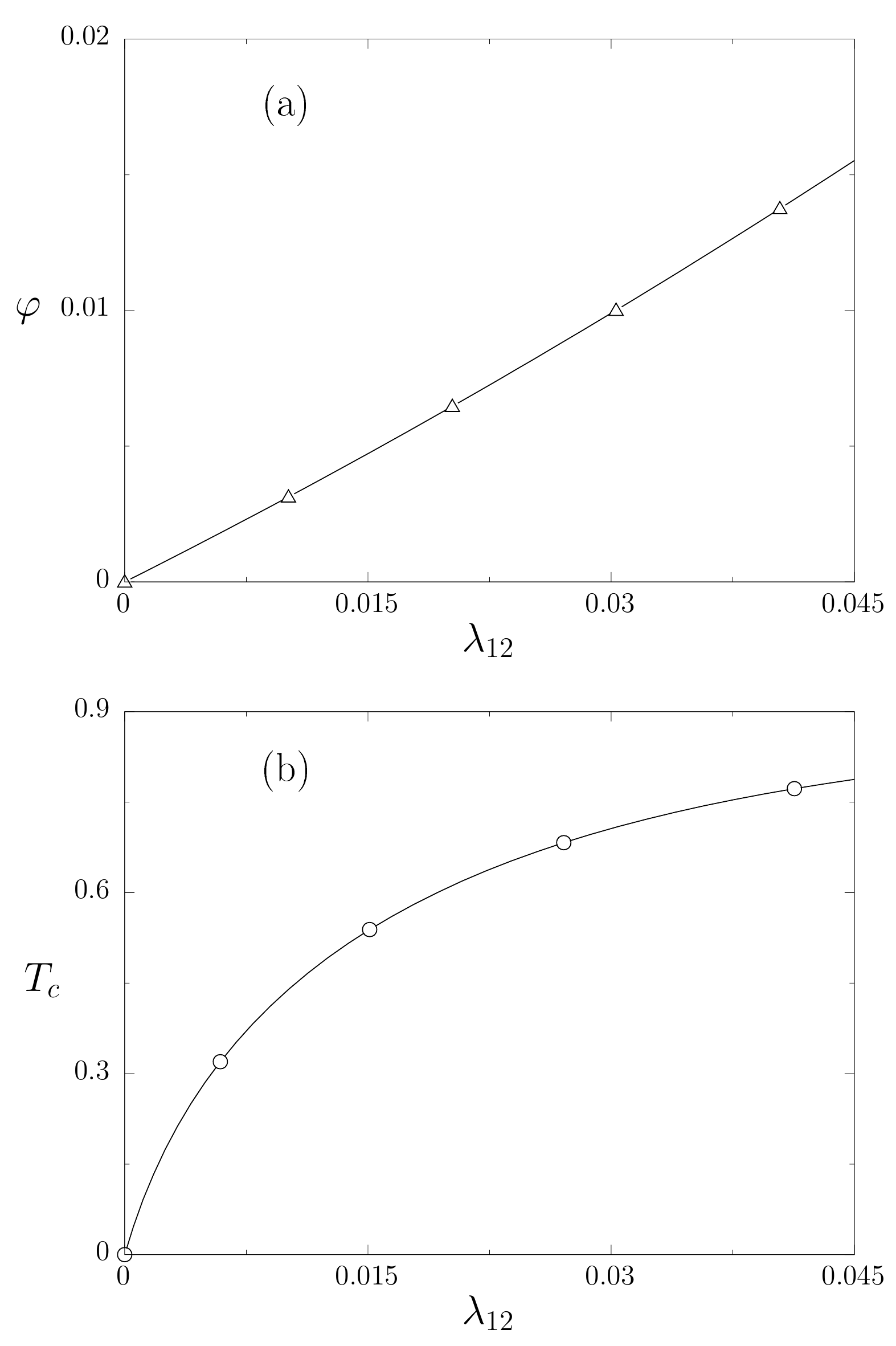} 
\caption{Numerical results obtained from  the effective potential, Eq.~(\ref{eq: num4.6}), in the case of  2d systems with $z=1$.   (a) The order parameter ($\varphi_1=\varphi_2=\varphi$) as a function of the effective coupling at zero temperature. One can confirm that quantum fluctuations enhance the coexistence region since $\varphi$ increases as a function of the quartic coupling $\lambda_{12}$. (b) The  critical temperature $T_c$ as a function of this coupling between the order parameters. These numerical results are consistent with the schematic phase diagrams exhibited in Fig.~\ref{fig1} and Fig.~\ref{fig3}.}\label{fig4}
\end{center}
\end{figure}

For completeness, and again, without loss of generality, we can obtain some numerical results using the same numerical parameters as before. For simplicity, we consider that  the two \textit{gaps} for thermal excitations are equal, see Eq.~(\ref{eq: num3.2}). This implies taking $\lambda_1=\lambda_2$, such that $\varphi_1=\varphi_2=\varphi$. Observe from Fig.~\ref{fig4}~(a) that, for zero temperature, the order parameter ($\varphi$) increases as a function of the couplings ($\lambda_{12}$), as expected since quantum fluctuations stabilize coexistence~\cite{35b}. Notice that $T_c$ also depends on the coupling $\lambda_{12}$ between the order parameters, as shown in Fig.~\ref{fig4}~(b).

\subsection{Finite temperature effects in coexistence case for two-dimensional case: Preamble}

The expression for the zero temperature effective potential for coexisting phases in 2d systems with $z=1$ in the presence of quantum corrections  is given by~\cite{35b},
\begin{align}
\begin{array}{ccc}
\small{V_{eff}}=\underbrace{r_1\varphi_{1}^{2}+\lambda_{1}\varphi_{1}^{4}+\lambda_{2}\varphi_{2}^{4}+\lambda_{12} \varphi_{1}^{2} \varphi_{2}^{2}+\delta_{1} \varphi_{1}^{3}\varphi_{2}+\delta_{2} \varphi_{1}\varphi_{2}^{3}}_{classical \ term} \\
-\frac{\sqrt{2}}{2 \pi}\Big[\frac{1}{3}\left((b_{1}+r_1)^{3/2}+b_{2}^{3/2}\right)-\frac{1}{3}\left|r_1\right|^{3/2} \\
\underbrace{+\frac{\left(3\delta_{1}\varphi_{1}^{2}+3\delta_{2}\varphi_{2}^{2}+4\lambda_{12}\varphi_{1}\varphi_{2}\right)^{2}}{4(\sqrt{b_{2}}+\sqrt{b_{1}+r_1})}}_{quantum \ corrections}\Big].
\end{array}
\label{eq: num4.7}
\end{align}

The system is fine-tuned at the ZTCCP of the phase characterized by $\varphi_2$, such that $r_2=0$. The quantities $b_{1,2}$ are given by Eq.~(\ref{eq: num3.2}). Notice that when $r_1 \rightarrow 0$ we recover the expression for the bicritical case, i.e., Eq.~(\ref{eq: num4.1}), as expected. 

We emphasize that the main difference between this case and the  bicritical one is that even at the classical level there is a coexistence region that is enhanced by quantum corrections for both couplings, $\lambda_{12}$ and $\delta_{1,2}$~\cite{35b}. In other words, both ZTCCPs move away due quantum fluctuations effects. Since the system is deep in the phase with $\varphi_1 \ne 0$  we can fix it at a constant (finite) value in order to analyze the temperature effects for this case, as we shall discuss later on. Moreover, we need to satisfy the condition $b_{1}>r_1$ in Eq.~(\ref{eq: num4.7}), otherwise we have domains formation where coexisting phases become metastable~\cite{35b,42} giving rise to non-homogeneous ground states.

\subsection{Finite temperature effects in coexistence case for two-dimensional case: Results}

Similarly to the section above for the bicritical case, we are interested in investigating finite temperature effects on the effective potential, Eq.~(\ref{eq: num3.5}), now for the coexistence case. In order to obtain some analytical expressions, we apply the same approximation from previous section, i.e., we also consider the low temperature regime, that is, for $T << \Delta_{1,2}$. In other words, again, one can neglect the two last temperature dependent terms in Eq.~(\ref{eq: num3.5}). However, the only difference is that in coexistence case the quantum correction term is given now by Eq.~(\ref{eq: num4.7}) from the previous section. 

Notice that for coexistence case, in the expressions $\epsilon_i$, Eq.~(\ref{eq: num3.5}), we take  $r_1 \neq 0$ and $r_2=0$. In this case we have the   following integrals to solve,
\begin{equation}
\int_0^{\infty} dp \ p \ \ln\left(1-e^{-\frac{\epsilon_{p_{i}}}{T}}\right) \approx T\left(\sqrt{b_{i}+r_{i}}+T\right)e^{-\frac{\sqrt{b_{i}+r_{i}}}{T}}.
\label{eq: num4.8}
\end{equation}
The integral in $\epsilon_{p2}$ has to be treated differently since $r_2=0$. Also,
\begin{equation}
\int_0^{\infty} dp \ p \ \ln\left(1-e^{-\frac{\epsilon_{p_{i}}}{T}}\right) \approx T\left(\sqrt{\left|r_{i}\right|}+T\right)e^{-\frac{\sqrt{\left|r_{i}\right|}}{T}}
\label{eq: num4.9}
\end{equation}
for $i \ne 2$. Using again that $\ln(1-x)\approx -x$, for small $x$, since we are interested in the low temperature regime $T << \Delta_{1,2}$.

Thus, we can rewrite the effective potential for coexistence case, taking into account finite temperature effects, with the help of the Eq.~(\ref{eq: num4.8}), Eq.~(\ref{eq: num4.9}) and Eq.~(\ref{eq: num4.7}),
\begin{align}
\begin{array}{cccc}
\small{V_{eff}}=\underbrace{r_1\varphi_{1}^{2}+\lambda_{1}\varphi_{1}^{4}+\lambda_{2}\varphi_{2}^{4}+\lambda_{12} \varphi_{1}^{2} \varphi_{2}^{2}+\delta_{1} \varphi_{1}^{3}\varphi_{2}+\delta_{2} \varphi_{1}\varphi_{2}^{3}}_{classical \ term} \\
-\frac{\sqrt{2}}{2 \pi}\Big[\frac{1}{3}\left((b_{1}+r_1)^{3/2}+b_{2}^{3/2}\right)-\frac{1}{3}\left|r_1\right|^{3/2} \\
\underbrace{+\frac{\left(3\delta_{1}\varphi_{1}^{2}+3\delta_{2}\varphi_{2}^{2}+4\lambda_{12}\varphi_{1}\varphi_{2}\right)^{2}}{4(\sqrt{b_{2}}+\sqrt{b_{1}+r_1})}}_{quantum \ corrections}\Big] \\
\frac{T^3}{(2\pi)^{2}}\Bigg\{-\sum_{i=1}^{2}\Bigg(\left[1+\frac{\sqrt{b_{i}+r_i}}{T}\right]e^{\left(-\frac{\sqrt{b_{i}+r_i}}{T}\right)}\\
\underbrace{-\left[1+\frac{\sqrt{\left|r_i\right|}}{T}\right]e^{-\frac{\sqrt{\left|r_i\right|}}{T}}\Bigg)}_{finite \ temperature \ effects}\Bigg\}
\end{array}
\label{eq: num4.10}
\end{align}
where $r_2=0$.

If $r_1=0$, we recover the result for bicritical case in the previous section, as expected (see Eq.~(\ref{eq: num4.6})). Differently from the  bicritical point discussed above, in coexistence case there is the emergence of a new characteristic length  associated with the \textit{mass} term ($r_1$)~\cite{10}. This can be seen directly from the solution of the integral in Eq.~(\ref{eq: num4.9}), i.e., the correlation length of the system is given by $\xi=1/\sqrt{2r_1}$, which is related to the distance to the ZTCCP~\cite{10}.

It's worth to point out that from the finite temperature terms in the effective potential, Eq.~(\ref{eq: num4.10}), we can recognize the same  temperature scaling of the free energy as in the bicritical case,  $f \propto T^{(d+z)/z}$,   for $T >>T_c$. Therefore, all the previous statements about the bicritical point, i.e., the \textit{weak first-order temperature phase transitions}~\cite{Pikin,Fernandez,Millis} and the specific heat regimes, hold even if we begin with a coexistence region in the classical phase diagram.

\begin{figure}[t]
\begin{center}
\includegraphics[width=1\columnwidth]{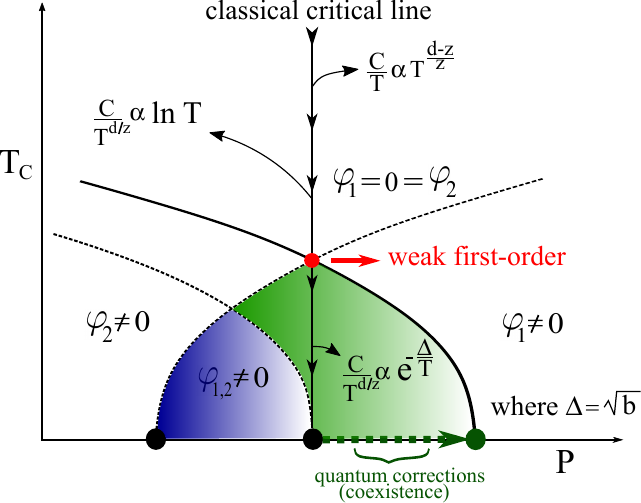} 
\caption{(Color online) Schematic phase diagram of the specific heat scaling regimes for coexistence region in 2d systems with $z=1$. Quantum corrections enhance the coexistence region (arrow (green)).  Fine tuning the external parameter $P$ at the  ZTCCP, we can identify three different scaling regimes for the specific heat. Notice that  in coexistence case there is the emergence of a new characteristic length of the system associated with the \textit{mass} term, i.e. $\xi=1/\sqrt{2r_1}$. The gaps for thermal excitations are now given by $\Delta_1=\sqrt{b_1+r_1}$ and $\Delta_2=\sqrt{b_2}$, where $b_{1,2}$ appear in Eq.~(\ref{eq: num3.2}). Analogously to the bicritical case, the point where both lines cross  (red) marks a \textit{weak first-order temperature phase transition} point. The intermediate regime, i.e., regime II, is distinguished by a logarithmic dependence of the specific heat. }\label{fig5}
\end{center}
\end{figure}

Analogously to the bicritical case, consider tuning the classical system to the ZTCCP of the phase characterized by $\varphi_2$, as in Fig.~\ref{fig5},  at th2 point $r_2=0$ of the phase diagram. At this point, we have $<\varphi_2>=0$, and  $\varphi_1=\left\langle \varphi_1\right\rangle=\sqrt{\left|r_1\right|/(2\lambda_1)}$ (classical value). As the quantum corrections are turned on, $<\varphi_2>$ becomes finite, due to the {\it repulsion} of the ZTCCPs that increases the region of coexistence~\cite{35b}. This order parameter $<\varphi_2>$, which was zero in the classical case now attains a finite value that  for a fixed point in the phase diagram inside the coexistence region increases, as the coupling $\lambda_{12}$ increases, see Fig.~\ref{fig5}. 

For numerical calculations we take the same previous parameters values. As temperature increases along the vertical line ($r_2=0, T \ne 0$) shown in Fig.~\ref{fig5}, the phase with $<\varphi_2>$ finite due to quantum corrections becomes unstable at a \textit{weak first-order transition}.
This is characterized by the minimum  of the temperature dependent effective potential at $<\varphi_2>=0$ becoming degenerate  with the minima at finite  $<\varphi_2> $.   

\begin{figure}[t]
\begin{center}
\includegraphics[width=1\columnwidth]{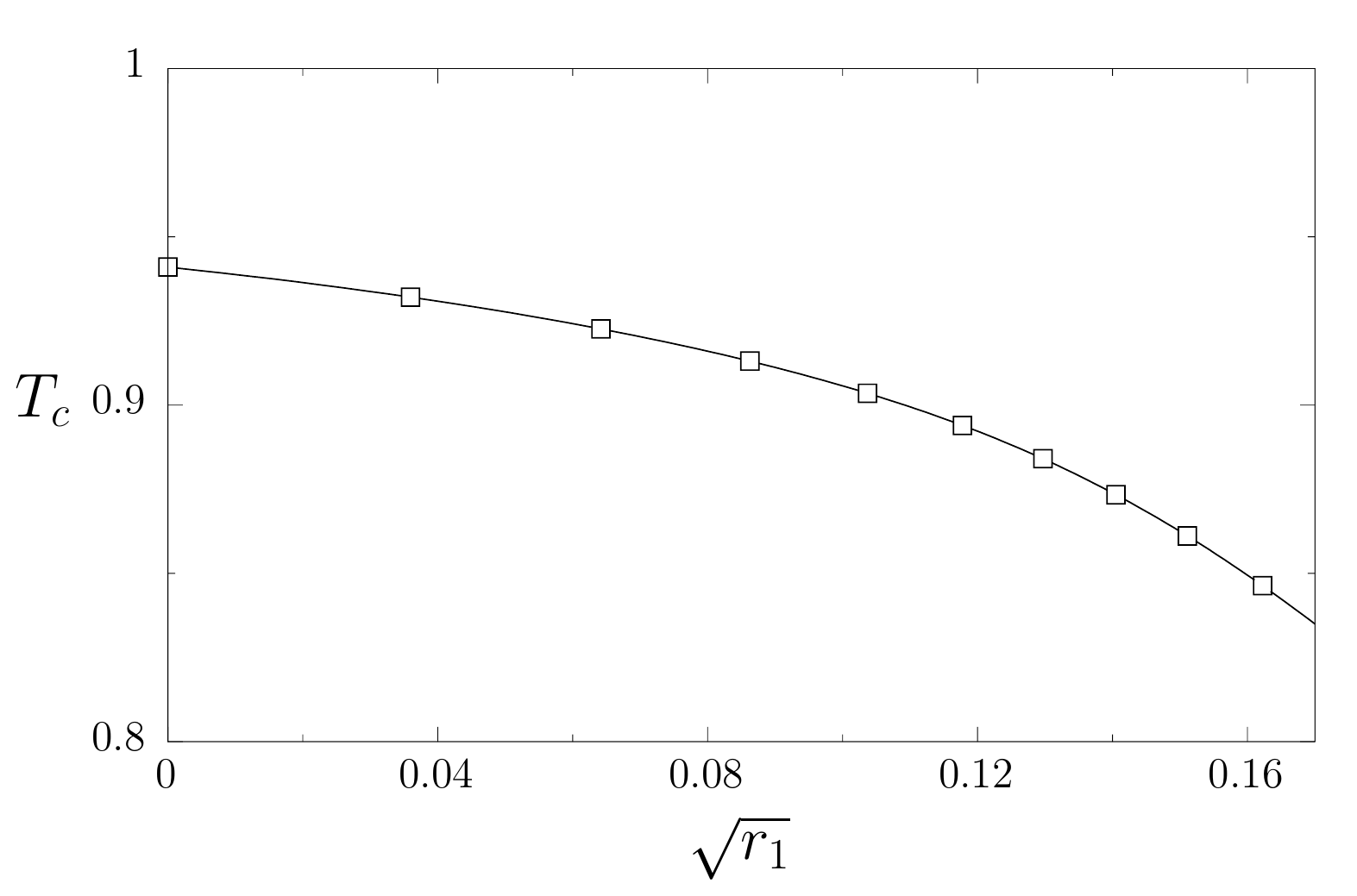} 
\caption{Numerical calculation of $T_c$ as a function of the distance to the ZTCCP ($r_1$) in the case of the 2d systems with $z=1$, i.e., using Eq.~(\ref{eq: num4.10}). $T_c$ decreases as we deviate from the ZTCCP. This  indicates that above where the ZTCCP is located, before including quantum corrections,  the maximum $T_c$ in the coexistence phase is attained. The numerical results are consistent with the schematic behavior presented in Fig.~\ref{fig5}.}\label{fig6}
\end{center}
\end{figure}

Finally, once we introduce thermal fluctuations, we can also investigate how $T_c$ behaves as a function of the distance to the ZTCCP in the coexistence region. This has  important experimental consequences for the region of coexistence in systems with competing orders. Note from Fig.~\ref{fig6} that since we deviate from the ZTCCP the critical temperature decreases. This suggests that above the fine tuned value of the ZTCCP, we may find the highest $T_c$ in the coexistence region~\cite{36,37,38,32,35,35b}. In other words, at finite temperatures, above the ZTCCP, the system presents the maximum $T_c$ of the coexistence region, as we consider both quantum  and finite temperature effects in the effective potential.

\subsection{Finite temperature effects in bicritical point and coexistence region for three-dimensional case}

As we have mentioned before, the extension to include finite temperature effects in 3d systems is straightforward. Using Eq.~(\ref{eq: num3.3}), Eq.~(\ref{eq: num3.5}), for 3d systems with a linear dispersion relation and for $T << \Delta_{1,2}$ the effective potential  becomes,
\begin{align}
\begin{array}{cc}
V_{eff} = V_{{eff}_{(T=0)}}^{(3+1)dim}\\
+\frac{\sqrt{2}}{(2 \pi)^3}2T^4\Bigg\{-\frac{1}{T^3}\int_0^{\infty} dp \ p^2 \ \left(e^{-\frac{\epsilon_{1}}{T}}+e^{-\frac{\epsilon_{3}}{T}}\right)\\
-\frac{1}{T^3}\int_0^{\infty} dp \ p^2 \ \left(e^{-\frac{\epsilon_{2}}{T}}+e^{-\frac{\epsilon_{4}}{T}}\right)\Bigg\}
\end{array}
\label{eq: num4.11}
\end{align}
where  $V_{{eff}_{(T=0)}}^{(3+1)dim}$ is the quantum corrections term at zero temperature previously obtained~\cite{35}, $\epsilon_{p_{1,3}}^{2}=p^2+r_{1,2}+b_{1,2}$, $\epsilon_{p_{2,4}}^{2}=p^2+r_{1,2}$, $\epsilon_{p_{5,6}}^{2} = p^2+(A,B)^2$, with $b_{1,2}$ and $(A,B)^2$ given in Eq.~(\ref{eq: num3.2}), and
\begin{equation}
f_{BE}(\epsilon_{p})=\frac{1}{e^{\epsilon_{p}/T}-1}
\label{eq: num26}
\end{equation}
is the Bose-Einstein distribution function~\cite{Matsubara,Kapusta,Abrikosov}.
In the case of a bicritical point, $r_1=r_2=0$, we use the result,
\begin{equation}
\int_0^{\infty} dp \ p^2 \ln \left( 1 - e^{-\frac{p}{T}} \right)= - \frac{1}{45} \pi^4 T^3
\end{equation}
to deal with the integration over $\epsilon_{2,4}$.

Notice the main differences when we compare  Eq.~(\ref{eq: num4.11}) to the effective potential for 2d systems, i.e., Eq.~(\ref{eq: num4.2}). In $V_{{eff}_{(T=0)}}^{(3+1)dim}$, the dependence of the integral on \textit{momentum},  Eq.~(\ref{eq: num4.11}),  is no longer linear but quadratic. The temperature dependent term in the effective potential now exhibits a $T^{4}$ dependence, instead of the $T^{3}$ in 2d. These results are consistent with the  $T^{(d+z)/z}$ dependence of the free energy expected from scaling close to a QCP~\cite{10}.
From the expression for $V_{{eff}_{(T=0)}}^{(3+1)dim}$ we find that in 3d systems, with $z=1$, the ZTCCPs are stable to quantum corrections in the presence of  an exclusive quartic interaction~\cite{35}.  However,  in the case of the bilinear coupling these corrections give rise to coexistence~\cite{35}, as in 2d systems~\cite{35b}.

The full results from Eq.~(\ref{eq: num4.11}) for both bicritical point and coexistence cases in 3d case have been obtained numerically. It turns out that the effects of thermal fluctuations in the case of bilinear coupling are very similar to those discussed above for the 2d case. Finite temperature effects, distinctively  from quantum fluctuations, tend to disorder the system as we increase temperature. As in 2d there are  two  regimes with different thermal behaviors related to the existence of a \textit{weak first-order transition}~\cite{Pikin,Fernandez,Millis} at $T_c$. In the high temperature regime, $T >> T_c$ the free energy  presents a temperature dependence of the form $T^{(d+z)/z}$ in agreement with the expected scaling behavior in the presence of a quantum critical point~\cite{10}. For the $T << T_c$ the specific heat is thermally activated with a gap related to the size of finite domains~\cite{10}.  

\section{Summary and conclusions} 
\label{sec:Conclusions}

Strongly correlated systems at low temperatures present complex phase diagrams~\cite{Braithwaite,Sundar,Balatsky1,Balatsky2,Aoki,Arce-Gamboa,1,Steglich,2,2.1,Park,4,5,6,7,8} with competing or coexisting orders that can be tuned by external parameters, such as, pressure, doping, or magnetic field~\cite{9,10}. The most notables are pnictides/iron-arsenide SC~\cite{11,12,13,14,15}, U and Ce-based heavy fermions~\cite{4,5,7} and High $T_c$ cuprates materials~\cite{4,5,7}, with unusual properties, due to competing/coexistence of different orderings. These may be structural transitions~\cite{16,19}, charge density waves, superconductivity and magnetism~\cite{Braithwaite,Sundar,Aoki}. All these effects are ongoing research topics, which need a fundamental theory to describe them.

In this paper, we  address the effects of finite temperature on our previous results concerning the stability of critical points on the ground state of the phase diagrams of systems with competing scalar orderings. We investigated the interplay between the effects of quantum and thermal fluctuations on the phase diagram of classical systems in two distinct conditions, i.e., bicritical point and coexistence case. The consideration of thermal fluctuations is essential if we want to make contact with experimental results. 

For simplicity, we considered two real scalar order parameters mediated by quartic as well as bilinear interacting terms for both three and two-dimensional systems described by a Lorentz invariant critical theory~\cite{35,35b}. In order to introduce finite temperature effects we have applied the well-known \textit{Matsubara summation} formalism from finite temperature quantum field theory~\cite{Matsubara,Kapusta,Abrikosov}.

We show that for two-dimensional systems, the ZTCBP is unstable for both quantum~\cite{35b} and thermal fluctuations in the cases of quartic and bilinear couplings of the order parameters. Increasing temperature gives rise to a \textit{weak first-order temperature phase transition}~\cite{Pikin,Fernandez,Millis} at $T_c$, where the coexisting phases exchange stability with a disordered phase. The \textit{weak first-order} nature of this transition is confirmed by a calculation of the latent heat that turns to be small compared with the thermal fluctuations at $T_c$. Also the observation of scaling behavior  of the free energy and specific heat above $T_c$ is consistent with the weak character of the transition~\cite{10}. This has as consequence that it is difficult to distinguish it from a continuous second order transition. 

Indeed, above $T_c$ the system will present scaling behavior associated with the existence of a QCP in d-dimensions and with dynamic exponent $z=1$ down to temperatures very close to $T_c$. In this region of the phase diagram for $T>>T_c$ the specific heat scales as $C/T \propto T^{(d-z)/z}$ that crosses over to a less universal behavior $C/T^{(d+z)/z} \propto- \ln T$ for $T \gtrsim T_c$ in both 2 and 3 dimensions. 

For low temperatures, $T <<T_c$ in the coexistence region,  we can identify the appearance of a gap for thermal excitations . This is related to  excitations in finite domains that nucleate below $T_c$ due to the first-order character of this transition~\cite{10}. Experimentally this gap manifests in a thermally activated contribution to the specific heat at these low temperatures. The length of the domains introduces new length scales in the problem~\cite{10}.

We have also studied the effects of thermal and quantum fluctuations in classical phase diagrams where there is a region of coexistence between two order parameters. In 2d  the zero temperature classical critical points are unstable to quantum fluctuations that enhance the region of coexistence for both quartic and bilinear couplings~\cite{35b}. At finite temperatures the coexisting phases exchange stability with a disordered phase at a \textit{weak first-order} transition. We find the maximum $T_c$ of a given phase occurs at temperatures above the ZTCCP of this phase. As before, scaling behavior is found for $T > T_c$ and thermally activated excitations for $T<< T_c$. 

In the case of bilinear coupling of the order parameters the results for 3d systems~\cite{35} are very similar to those in 2d~\cite{35b}. We have to remark the stability of the zero temperature bicritical point to a quartic coupling in 3d~\cite{35}. In the case of coexistence, quantum fluctuations in the presence of a quartic coupling in 3d actually {\it reduce} the region of coexistence. In this sense, it competes  with the bilinear interaction~\cite{35} that acts to increase this region.

From numerical calculations we notice that in 3d,  thermal fluctuations, in opposite to quantum fluctuations, present some common features with the 2d case. In both, 2d and 3d,  thermal fluctuations lead to disorder in the system and give rise to \textit{weak-first order transitions} as we increase temperature. Above this transition, consistent with its weak nature we can identify scaling behavior as approaching a QCP.  

Finally, our study clearly points out the ubiquity of \textit{weak first-order transitions} in 2d systems with competing order parameters.
Although our study has been carried out for QCPs with Lorentz invariance~\cite{36,37,38,32,35,35b}, we expect that our general conclusions as the existence of these finite temperature transitions and the accompanying scaling behavior will persist for arbitrary values of the dynamic exponent.

\section{acknowledgments}
The Brazilian agencies {\em Conselho Nacional de Desenvolvimento Cient\'\i fico e Tecnol\'ogico} (CNPq), {\em Funda\c c\~ao  Carlos Chagas Filho de Amparo \`a Pesquisa do Estado do Rio de Janeiro} (FAPERJ), and {\em Coordena\c c\~ao de Aperfei\c coamento de Pessoal de N\'\i vel Superior} (CAPES) are acknowledged for partial financial support.

\end{document}